\documentclass[10pt]{article}
\usepackage{epsfig,dsfont}

\begin{document}
\sffamily

\thispagestyle{empty}
\vspace*{15mm}

\begin{center}

{\LARGE 
Canonical fermion determinants in lattice QCD -- 
\vskip3mm
Numerical evaluation and properties}
\vskip15mm
Erek Bilgici$^a$, Julia Danzer$^a$, 
Christof Gattringer$^a$, C.B.~Lang$^a$, and Ludovit Liptak$^b$
\vskip8mm
$^a\,$Institut f\"ur Physik, Unversit\"at Graz, \\
Universit\"atsplatz 5, 8010 Graz, Austria 
\vskip5mm
$^b\,$Institute of Physics, Slovak Academy of Sciences, \\
Dubravska cesta 9, 845 11 Bratislava, Slovak Republic 

\end{center}
\vskip30mm

\begin{abstract}
We analyze canonical fermion determinants, i.e., fermion determinants  projected
to a fixed quark number $q$. The canonical determinants are computed  using a
dimensional reduction formula and are studied for pure SU(3)  gauge
configurations in a wide range of temperatures. It is demonstrated that the
center sectors of the Polyakov loop very strongly manifest themselves in the
behavior  of the canonical determinants in the deconfined phase, and we discuss
physical implications of this finding. Furthermore the distribution of the quark
sectors is studied as a function of the temperature.
\end{abstract}

\setcounter{page}0
\newpage
\noindent
{\large \bf Introductory remarks}
\vskip3mm
\noindent
With the  heavy ion experiments at RHIC, LHC and FAIR, Quantum
Chromodynamics with temperature and non-vanishing density is currently a 
focus of attention. Of particular interest are non-perturbative results for
several questions, such as  obtaining information about the phase diagram, 
characterization of the different phases, and a possible understanding of
the mechanisms that drive the various transitions. In principle lattice QCD
allows for such a non-perturbative approach, but numerical simulations at finite
density are plagued by serious phase cancellation problems. 

Recently the canonical approach, where one works with a fixed net quark number,
was addressed in various papers \cite{canonical1} -- \cite{review}. The connection
between the conventional grand canonical determinant with a chemical potential 
and the canonical approach can be made with a fugacity expansion where the
expansion coefficients are the  canonical determinants, i.e., fermion
determinants projected to a  fixed net quark number. Using a recently proposed
dimensional reduction for the fermion determinant \cite{DaGa} we can numerically
evaluate canonical determinants efficiently and study their properties. 

Canonical determinants are not only objects that might provide new and more
efficient strategies for simulating lattice QCD with finite density, but also have
interesting properties themselves. In particular they  transform  in a simple
way under center rotations -- a symmetry that is of crucial
importance for understanding the high temperature phase transition of pure gauge
theory. In this article we analyze the influence of center symmetry
and its breaking for the canonical 
fermion determinants and study the distribution of the
quark number as a function of temperature. 
 
\vskip5mm
\noindent
{\large \bf Canonical fermion determinants and dimensional reduction}
\vskip3mm
\noindent
In this article we work with Wilson's lattice Dirac operator $D(\mu)$ with
chemical  potential $\mu$ on lattices of size $L^3 \times \beta$, where $\beta$
is the  inverse temperature in lattice units, i.e., the number of lattice sites
in time direction (see \cite{DaGa} for the details of our conventions). The grand
canonical partition sum $Z_{gc}$ is obtained as the path integral over gluon
($U$) and quark ($q,\overline{q}$) degress of freedom,
\begin{equation}
Z_{gc} \; = \; \int \!\! {\cal D} [U,q,\overline{q}] \, 
e^{\, - \, S_G[U] \, - \, S_F[U,q,\overline{q}]} \; = \; 
\int \!\! {\cal D} [U] \, 
e^{\, - \, S_G[U]} \, \det[D(\mu)]_{gc}^{\; N_f} \; ,
\end{equation}
where $S_G$ and $S_F$ denote the gauge and fermion (including
$\mu$) parts of the action. In the
second step we have integrated out the $N_f$ flavors of quarks and 
obtained the grand canonical fermion determinant $\det[D(\mu)]_{gc}$.  
The grand canonical fermion determinant can be represented as a 
fugacity series, 
\begin{equation}
\det[D(\mu)]_{gc} \; = \; \sum_q e^{q \mu\beta} \, \det[D]^{(q)} \; ,
\label{detfugacity}
\end{equation}
where the sum runs over integer valued quark numbers $q \in [-6L^3,+6L^3]$. 
The expansion coefficients $\det[D]^{(q)}$ are the 
canonical determinants and may be obtained using Fourier transformation
with respect to an imaginary chemical potential,
\begin{equation}
{\det}[D]^{(q)} \; = \; \frac{1}{2\pi} \int_{-\pi}^\pi \!\!\!\! d \varphi  \;
e^{-i q \varphi}\, \det[D(\mu = i\varphi/\beta)]_{gc} \;\; .
\label{DQfourierdef} 
\end{equation} 
Exploring the generalized 
$\gamma_5$-hermiticity, $\gamma_5 D(\mu) \gamma_5 = D(-\mu)^\dagger$,  
one finds the relation $\det[D]^{(-q)} \, = \, ( \det[D]^{(q)} )^*$
between canonical determinants with positive and negative quark numbers
$q$. For vanishing quark number  $q = 0$ the canonical determinant is real, 
i.e., ${\det}[D]^{(0)} \in \mathds{R}$.

The grand canonical determinant $\det[D(\mu)]_{gc}$ is a gauge invariant object
and as such is built from products of closed loops of gauge links on the
lattice. The projection (\ref{DQfourierdef}) to a fixed quark number $q$ selects
the  subset of loops with a net winding number of $q$. This implies that the 
canonical determinant $\det[D]^{(q)}$ is made from loops with a net winding 
number of $q$ and thus transforms in a simple way under center rotations,
where at a fixed time slice $x_4^*$ all temporal links  are multiplied by a
fixed  element $z$ of the center $\mathds{Z}_3 =  \{ 1 \, , \, e^{i2\pi/3} \, ,
\, e^{-i2\pi/3} \}$ of the gauge group, i.e.,  $U_4(\vec{x},x_4^*) \;
\longrightarrow \; z \,U_4(\vec{x},x_4^*)$. The transformation law is
\begin{equation}
{\det}[D]^{(q)} \; \longrightarrow \; z^q \, {\det}[D]^{(q)} \; = \; 
z^{q\,\mbox{\textrm{\scriptsize mod}}\,3} \, {\det}[D]^{(q)} \; ,
\label{DQcentertrafo}
\end{equation}
i.e., $\det[D]^{(q)}$ picks up a phase which depends on the quark number $q$.

Using a domain decomposition technique it was shown \cite{DaGa}  that the
grand canonical fermion determinant may be rewritten exactly in a dimensionally
reduced form,
\begin{equation}
\det[D(\mu)]_{gc} \; = \; A_0 \; \det 
\big[ \mathds{1} - H_0 - e^{\mu \beta} H_{+} - e^{-\mu \beta} H_{+}^{\, \dagger} \big] \; .
\label{detH0Hpm}
\end{equation}
Here $A_0$ is a real factor which depends only on the background gauge field
configuration  but is independent of the chemical potential $\mu$. $H_0 =
H_0^\dagger$,  and  $H_{+}$ are matrices that are built from propagators on  the
domains of the lattice and live on only a single time slice  (see \cite{DaGa}
for details). Thus the determinant in (\ref{detH0Hpm}) is dimensionally reduced,
i.e., the determinant is taken over a  matrix with $12 V$  rows and columns,
where $V$  is the spatial volume and the factor 12  comes from the color and
Dirac indices. We will explore  the dimensional reduction formula
(\ref{detH0Hpm}) to speed up the numerical evaluation  of the canonical
determinants with the Fourier integral (\ref{DQfourierdef}). We remark at this
point that recently \cite{altfactwilson} 
a complete factorization of the $\mu$-dependence of the determinant 
was presented also for the case of
Wilson fermions -- for staggered fermions such a result has been known before
\cite{altfact}. After the complete 
diagonalization of a reduced matrix this may be used to
evaluate the determinant for arbitrary many values of $\varphi$ and a very
efficient determination of the Fourier integral (\ref{DQfourierdef})
with high precision may become possible.  
 
\vskip5mm 
\noindent
{\large \bf Setting of the numerical analysis}
\vskip3mm
\noindent 
In this letter we study numerically the properties of the canonical
determinants  $\det[D]^{(q)}$ using quenched finite temperature
configurations on $6^3\times 4, 8^3 \times 4$ and $10^3 \times 4$
lattices. We work with the L\"uscher-Weisz gauge action
\cite{luweact} and for an estimate of the  scale  we in some places use the
lattice spacing determined in \cite{scale} from the Sommer parameter. Our
temperatures range between $0.7\, T_c$ and $1.43 \, T_c$, where we use the
critical temperature determined in \cite{tcrit}. For the update we 
use a mix of overrelaxation and Metropolis sweeps combined with a random 
center rotation to update this symmetry of the quenched theory. The bare mass
parameter of the Dirac operator was set to $m = 100$ MeV.  All errors we show
are statistical errors determined with single elimination Jackknife.

The canonical determinants $\det[D]^{(q)}$ were computed by numerically 
evaluating the Fourier representation (\ref{DQfourierdef}). In the Fourier
integral we use the representation (\ref{detH0Hpm}), which has the advantage
that the $\mu$-dependent part is a determinant which is dimensionally
reduced and thus allows for a numerical evaluation which is of ${\cal
O}(\beta^3) = {\cal O}(64)$  times faster (the terms $H_0$ and $H_+$ were
pre-calculated and stored in memory).  The numerical evaluation of the
Fourier sums was done with 64 values of $\varphi$ in the interval
$[-\pi,+\pi]$ using interpolation techniques. The logarithm of
$\det[D]_{gc}$ exhibits a smooth behavior. We thus first fitted it with
local cubic spline functions. The exponential of this interpolating function
was then used as weight in the integral (\ref{DQfourierdef}) leading to the
Fourier coefficients. The integral was performed with the adaptive 
trapezoidal rule. In various tests this approach turned out to be the most
stable and reliable one.

The correct implementation and accuracy of the evaluation of the canonical 
determinants can be checked by comparing the fugacity expansion
(\ref{detfugacity}) with a direct evaluation of $\det [D(\mu)]$ for
individual gauge configurations. This test was implemented using canonical 
determinants ${\det}[D]^{(q)}$ with quark numbers up to $q = 10$ in the 
fugacity expansion. With a cut of $q = 10$, for most of the configurations 
one can reproduce $\det [D(\mu)]$  with very good
accuracy for a large range of parameter values: For temperatures up  to $T_c$ the
relative error typically ranges from $10^{-7}$ to $10^{-2}$ for 
$a\mu \in [0,0.2]$,
with the best results for smaller $\mu$ and the lower temperatures.
This behavior is to be expected, since both, increasing the temperature or
the chemical potential, populates  higher quark numbers $q$ and terms from
$q$-values larger than  our maximal value of $q = 10$ become important. For
that reason, above $T_c$  the fugacity expansion with only terms up to $q =
10$  reproduces the grand canonical determinant only with poor accuracy. We
stress  that this is not a problem in principle, since for different
(dynamical)  configurations (results will be reported elsewhere),  where we
worked with 256 intermediate values in the  Fourier integral and used terms
up to $q = 40$, the fugacity expansion  reproduces the grand canonical
determinant very accurately also in  the deconfined phase. 

We remark at this point, that we experimented also with a perturbative approach
\cite{DaGa} for computing the canonical determinants ${\det}[D]^{(q)}$ based on 
(\ref{detH0Hpm}), and an alternative Fourier-based expansion suggested in 
\cite{kentuckymethod}. Both expansions necessarily have to be cut at some order,
and we found that for larger values of $q$, where the cut has to be set  to
higher values, the two expansion methods are not competitive with the Fourier
method combined with dimensional reduction. The problem  is most severe for high
temperatures and larger volumes where contributions  from high quark numbers $q$
are important.

\begin{figure}[p]
\begin{center}
\includegraphics[height=73mm,clip]{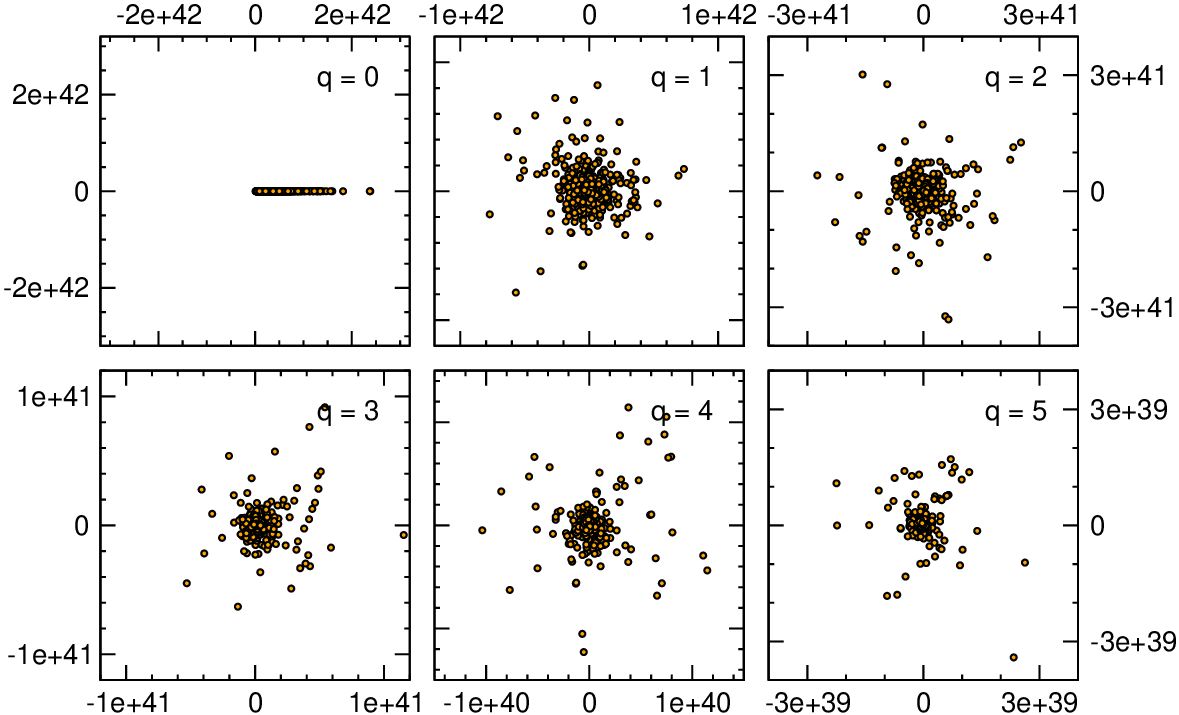}
\vskip10mm
\includegraphics[height=73mm,clip]{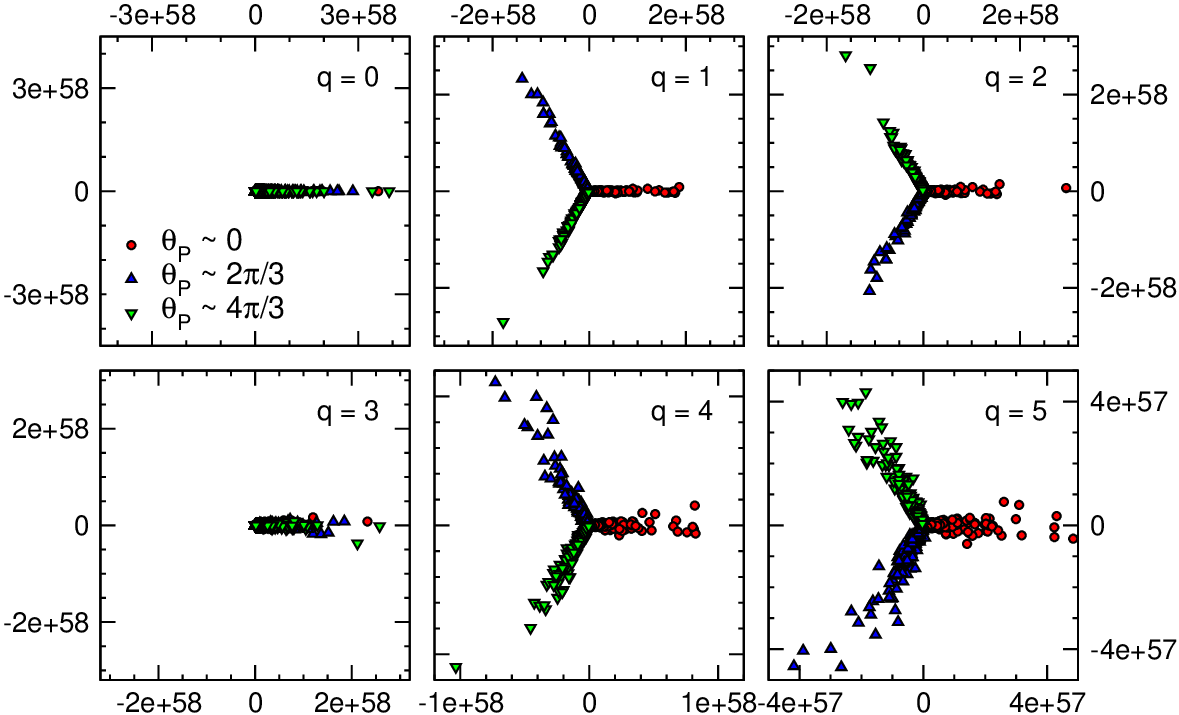}
\end{center}
\caption{ Scatter plots in the complex plane for the canonical
  fermion determinants $\det[D]^{(q)}$  for $q = 0,1,2 \, ... \, 5$  in the
  center symmetric phase (top set of plots, $T = 0.7 \, T_c$) and the center broken 
  phase (bottom, $T = 1.43 \, T_c$).  The data are from 1000 configurations on our $8^3
  \times 4$ lattices, and for the $T > T_c$ data we also indicate to which center sector 
  the underlying 
  gauge configuration belongs; we use the phase $\theta_P$ of the Polyakov
  loop to define the sector.  
\label{DQscatter}}
\end{figure}

\newpage
\noindent
{\large \bf Center properties of the canonical determinants}
\vskip3mm
\noindent
We begin our analysis of the canonical determinants by showing in
Fig.~\ref{DQscatter} scatter plots of the values of  $\det[D]^{(q)}$ in the
complex plane. We compare $q = 0,1, \, ... \,5$ on $8^3 \times 4$ lattices
at two different values of the gauge coupling, giving rise to an ensemble in
the center symmetric phase and one above $T_c$.  The statistics is 1000
configurations for  both temperatures, and in the $T > T_c$ data points we
distinguish the phase $\theta_P$ of the Polyakov loop of the underlying
configuration. Above $T_c$ the Polyakov loop shows the
well known pronounced concentration near the center phases  $e^{i \theta_P}
\sim 1, e^{i 2\pi/3}$ or $e^{-i 2\pi/3}$. We refer to the three subsets of
gauge  configurations with Polyakov loop phases $e^{i\theta_P}$  near these
three values as the real and the  complex center sectors. 

The plots demonstrate that the canonical determinants clearly distinguish 
between the center symmetric phase below $T_c$ (top set of plots in
Fig.~\ref{DQscatter})  and the center broken phase (bottom, $T > T_c$). In the
center symmetric phase the values of the canonical determinants scatter
isotropically around the origin and the absolute  values are small compared to
the ones found above $T_c$. In the center  broken phase for $q\,$mod$\,3 \neq 0$
we observe the characteristic center pattern  familiar from the Polyakov loop.
This behavior may be understood  as a consequence of the transformation
properties (\ref{DQcentertrafo}) of ${\det}[D]^{(q)}$ under center
rotations, which either transform trivially  (if $q\,$mod$\,3 = 0$), in the
same way as the Polyakov (if $q\,$mod$\,3 = 1$), or as the conjugate
Polyakov loop (if $q\,$mod$\,3 = 2$). Thus one may expect that the canonical
determinants ${\det}[D]^{(q)}$ with  $q\,$mod$\,3 \neq 0$ behave similar to
the Polyakov loop and the plots in  Fig.~\ref{DQscatter} confirm this
expectation (note the information on the center  sectors of the underlying
gauge configuration which we encode in the different symbols used for $T >
T_c$).  We stress that the center pattern  above $T_c$ for small values of
$q$ is much sharper than for the Polyakov loop on a lattice of same size.
This is due to the fact, that many different loops and  their products with
total winding number $q$ contribute to ${\det}[D]^{(q)}$ and their
self-averaging makes the center pattern more pronounced. A similar effect
was  found for the dual chiral condensate which also consists of sets of
loops with a common net winding number \cite{dualcond}. We stress at this point 
that below $T_c$, where the center symmetry is unbroken, the transformation law 
(\ref{DQcentertrafo}) implies that ${\det}[D]^{(q)} = 0$, unless 
$q\,$mod$\,3 = 0$. For canonical determinants with vanishing 
triality a non-isotropy in the distribution -- which is however hard to spot in 
the scatter plots -- can give rise to a non-zero value also below $T_c$.
This property reflects the fact that the baryon number has to be integer and
fractional baryon number, i.e., colored degrees of freedom, may emerge only in the
high temperature phase where the center symmetry is spontaneously broken.

Let us now discuss an important consequence of the strong center pattern
observed in the canonical determinants. The transition from the real center
sector to the sector with $\theta_P \simeq 2\pi/3$ may be implemented by a
center rotation of the gauge configuration with $z = \exp(i 2 \pi/3)  \sim
\exp(i \theta_P)$, and equivalently for the other complex center sector.
According to Eq.~(\ref{DQcentertrafo}) the canonical determinants 
transform as $\det[D]^{(q)}
\rightarrow z^q \det[D]^{(q)}$ under
this rotation. Using this in the  fugacity expansion
(\ref{detfugacity}) we obtain for the grand canonical
determinant in the different center sectors,
\begin{equation}
\det[D(\mu)]_{gc} \, \big \vert_z \; = \; \sum_q z^q \, {\det}[D]^{(q)} \, e^{\mu q \beta} \; .
\label{fugexppol}
\end{equation}
Exploring this equation we now discuss that in the high
temperature phase the grand canonical determinant must
behave quite differently in the different center sectors. 
Let us first consider the real center
sector where $z = 1$. Then the fugacity
expansion (\ref{fugexppol}) is a sum of
essentially real and positive terms. The fact that
for $\theta_P = 0$ the $\det[D]^{(q)}$ are essentially real and
positive follows from the scatter plots in
Fig.~\ref{DQscatter}.   

\begin{table}[b]
\begin{center}
\begin{tabular}{ccc}
\hline
\hline
\qquad $\;a \mu$ \qquad\qquad  & \qquad $\theta_P \simeq 0$ 
\qquad \qquad& \qquad $\theta_P \sim \pm 2\pi/3$  \qquad\qquad \\         
\hline
  0.00  &  $0.175(15) \times 10^{59}$ &  $0.45(13) \times 10^{53}$  \\
  0.05  &  $0.212(19) \times 10^{59}$ &  $0.43(14) \times 10^{53}$  \\
  0.10  &  $0.379(36) \times 10^{59}$ &  $0.37(14) \times 10^{53}$  \\
  0.20  &  $0.447(51) \times 10^{60}$ &  $0.26(17) \times 10^{53}$  \\   
  0.40  &  $0.109(27) \times 10^{66}$ &  $0.47(47) \times 10^{53}$  \\
  0.60  &  $0.48(27)  \times 10^{79}$ &  $0.11(11) \times 10^{55}$  \\
\hline
\hline
\end{tabular}
\end{center}
\caption{Results for $\langle | \det[D(\mu)] | \rangle_G$ in the real and
complex center sectors
at various values of the chemical potential $\mu$  
for our $8^3 \times 4$, $T = 1.43 \, T_c$ ensemble.} 
\label{gcdet_ploopsector}
\end{table}  

The complex center sectors, on the other hand,  
are characterized by $\theta_P \simeq \pm 2\pi/3$. As discussed in the 
last paragraph, these phases 
can also be obtained from the real sector by a center rotation with 
$z = \exp(\pm i 2\pi/3)$. Inserting these values of 
$z$ into (\ref{fugexppol})
leads to relative complex phases in the fugacity expansion and thus to
cancellations. Consequently one expects that the
values of the grand canonical determinant are smaller
for the complex Polyakov loop sectors.

In order to study this effect we analyzed the
expectation value  of the modulus of the grand
canonical determinant, $\langle | \det[D(\mu)] |
\rangle_G$, again dividing the gauge
configurations into the three center
sectors. $\langle .. \rangle_G$ denotes the 
expectation value of pure gauge theory.
Table \ref{gcdet_ploopsector} shows the
corresponding results for $T = 1.43 \, T_c$. 
It is obvious from the table, that for
the complex center sectors the average size of
the grand canonical determinant is several orders of magnitude 
smaller than for the
real center sector. Thus the latter sector
receives a much larger weight in the path integral
and thus is selected by the system in the high
temperature phase. 
In other words, in full QCD the pure gauge configurations are weighted with the
determinants and thus the $\theta_P = \pm 2\pi/3$ sectors are suppressed. 
When analyzing $\langle | \det[D(\mu)] |
\rangle_G$ at low temperature, we found
essentially no discrepancy between the three center
sectors. We conclude that in the high temperature phase
the selection of the
real center sector in the dynamical case   
can be understood as a consequence of the center 
symmetry properties of the canonical determinants.

\newpage
\noindent
{\large \bf Size distribution properties of canonical determinants}
\vskip3mm
\noindent
Analyzing scatter plots we established interesting properties for the phase
distribution of the canonical determinants at high temperature, which, as we
illustrated,  is related to center symmetry. Now we demonstrate that also
the modulus of the canonical determinants shows an interesting behavior. 

\begin{figure}[t!]
\begin{center}
\hspace*{-7mm}
\includegraphics[height=59mm,clip]{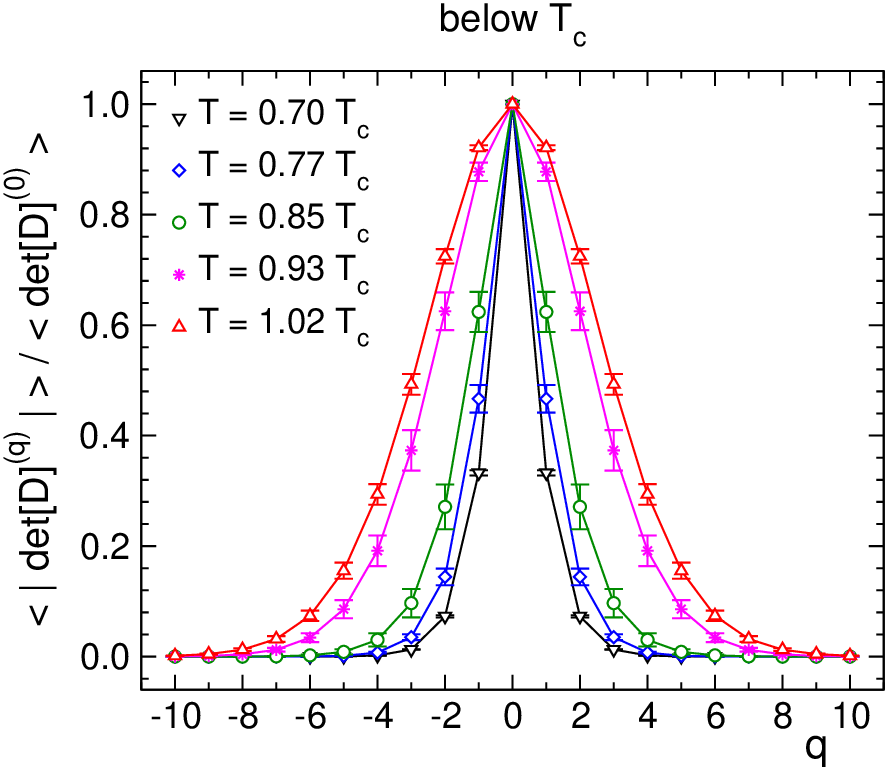}
\hspace{-2mm}
\includegraphics[height=59mm,clip]{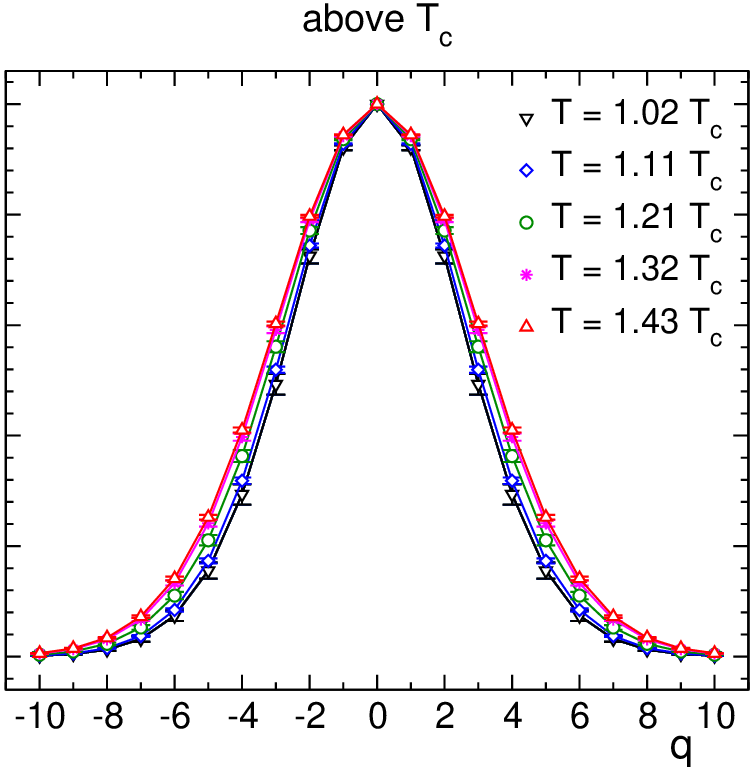}
\end{center} 
\caption{The average modulus of the determinant normalized relative to the $q=0$ 
case, $\langle | \det[D]^{(q)} | \rangle /  \langle \det[D]^{(0)} \rangle$, 
as a function of $q$. In the lhs.\ plot we show the results   
for temperatures up to $T_c$, and in the rhs.\ plot the 
results for temperatures above $T_c$. As we discuss in the text, the width of the
distribution is related to the quark number 
susceptibility evaluated at $\mu = 0$.
The data are from our $8^3\times 4$ ensembles.}
\label{DQratiovev}
\end{figure}

We begin our discussion with Fig.~\ref{DQratiovev}, where we plot the average
over the modulus of the canonical determinants 
$\langle | \det[D]^{(q)} | \rangle  /  \langle \det[D]^{(0)} \rangle_G$
as a function of $q$, normalized
relative to the $q=0$ determinant (which is real). 
In other words we study the average size of
the canonical determinant for different $q$. In the lhs.\ panel of
Fig.~\ref{DQratiovev} we show for the $8^3 \times 4$ ensembles the results in
the low temperature phase, while the rhs.\ is for high temperature.
The distributions are
symmetric around $q = 0$
(as they must be since $\det[D]^{(-q)} = (\det[D]^{(q)})^*$) and show a
Gaussian-like behavior as a function of $q$. The distribution is rather
narrow for the lowest temperature ($T = 0.7 \, T_c$) and widens as 
the temperature is increased. Above $T_c$ (rhs.\ panel) the widening of the 
distribution with increasing $T$ saturates, and the width of the Gaussian 
remains almost constant above $T_c$.

\begin{figure}[t]
\begin{center}
\includegraphics[width=80mm,clip]{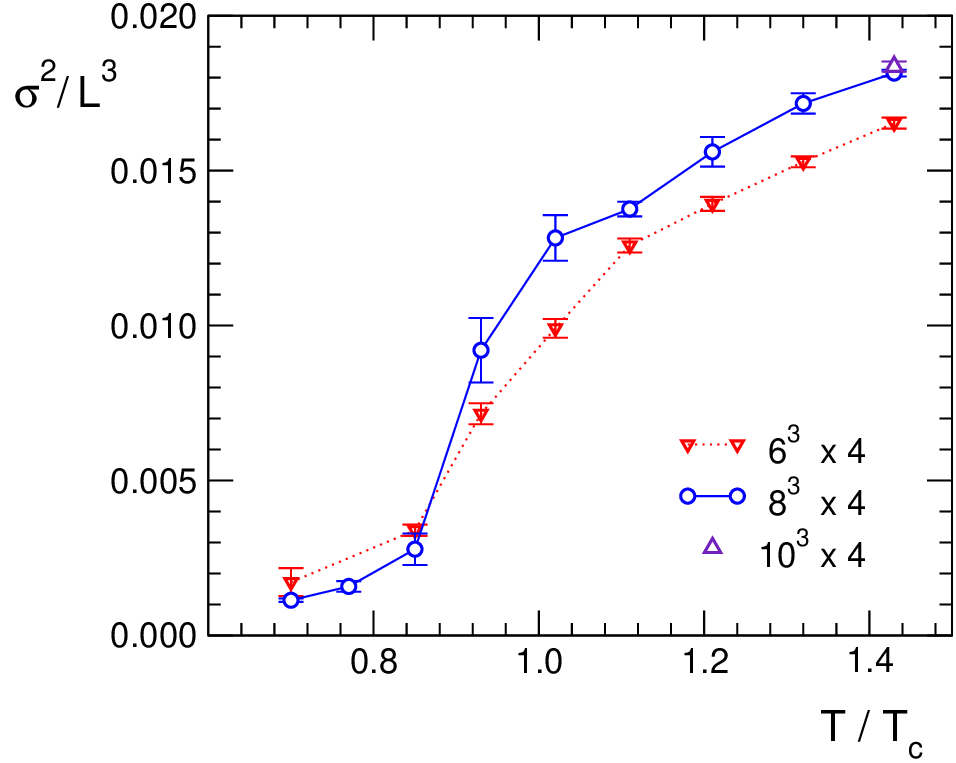}
\end{center}
\vspace{-5mm}
\caption{The square of the width of the distribution of the $\det[D]^{(q)}$ 
normalized by the spatial volume, $\sigma^2 / L^3$, as a function of
temperature.}
\label{width_vs_T}
\end{figure}

In order to analyze the behavior of the width of the distribution of
$\det[D]^{(q)}$, we fit the data of Fig.~\ref{DQratiovev} with a Gaussian
$\exp(-q^2/2\sigma^2)$ where the width $\sigma$ is the fit parameter. 
Fig.~\ref{width_vs_T} shows how $\sigma^2/L^3$, i.e.,
the width squared normalized by the spatial volume behaves as a function of
$T$. The data show a strong increase of $\sigma^2/L^3$ near the
critial temperature $T_c$. The increase is more pronounced for the 
$8^3 \times 4$ lattice and smoother for the $6^3 \times 4$ data --
the well known rounding of critical behavior for finite volume. Our 
$10^3 \times 4$ result agrees well with the $8^3 \times 4$ data. 
 
Let us now attempt an interpretation of the width $\sigma$ using a simple model: 
In Fig.~\ref{DQscatter} we found that for gauge configurations in the real
center sector the results for the canonical determinants $\det[D]^{(q)}$ are 
either very small (below $T_c$) or are close to the positive real axis (above $T_c$), such that 
we can ignore the phases of the canonical determinants $\det[D]^{(q)}$ and identify
$\langle \det[D]^{(q)} \rangle_G \sim \langle |\det[D]^{(q)}| \rangle_G$. 
Furthermore, as we have illustrated 
in Table 1, the contribution of the complex center sectors 
is highly suppressed. Using the Gaussian from the last paragraph we  
find 
$\langle \det[D]^{(q)} \rangle_G \sim \langle \det[D]^{(0)} \rangle_G 
\exp(-q^2/2\sigma^2)$. Combining this with the 
fugacity expansion (\ref{detfugacity}) and  
$Z(\mu) = \langle \det[D(\mu)] \rangle_G$ 
for the grand canonical partition sum, we find 
$\chi_q \, = \, \beta^{-2} \partial/ \partial \mu^2 \, \ln Z \, = \, \sigma^2$. 
In other words, the width of the distribution of the $\det[D]^{(q)}$ is related to
the quark number susceptibility $\chi_q$ evaluated at $\mu = 0$, 
and Fig.~\ref{width_vs_T} shows that 
$\sigma$ indeed behaves as one expects for $\chi_q$. In particular 
$\chi_q$ is an extensive quantity and the results for $\chi_q/L^3 = \sigma^2/L^3$
from different volumes should agree. Given the small volumes we can work with, Fig.~\ref{DQscatter}
illustrates this behavior rather well. This interpretation of 
$\sigma^2 \sim \chi_q$ of course ignores the back-reaction of the fermions which
is absent in our quenched 
gauge ensembles (compare the concluding discussion).
 
\vskip5mm
\noindent
{\large \bf Summary and outlook}
\vskip3mm
\noindent 
In this article we discuss the numerical evaluation and properties of
canonical fermion determinants using Wilson fermions on pure SU(3)
configurations at finite temperature. 
We obtain a considerable speedup in the evaluation of the
canonical determinants by implementing a domain decomposition approach which
leads to a dimensional reduction of the fermion determinant \cite{DaGa}.

We evaluate the canonical determinants with quark numbers
$q \in [-10,10]$, which
for our lower temperatures and not too large chemical potential is sufficient
for a very accurate representation of 
the grand canonical determinant through the
corresponding fugacity expansion. The obtained accuracy of the fugacity
expansion is an important test for the correctness of our implementation. For
larger chemical potential and high temperatures larger values of the quark
number $q$ would be needed in the fugacity representation. 

We then explore the physical properties of the 
canonical determinants. Scatter
plots of the canonical determinants show that at low temperature the
determinant values are small compared to high temperatures and that
they  are distributed 
isotropically in the complex plane.  
In the high temperature phase we observe a pronounced center pattern for
the canonical determinants and show that this pattern behaves exactly as
expected from the center transformation properties of the canonical
determinants. 
Based on the observation that the center symmetry is so strongly
manifest in the canonical determinants, we explain and
numerically confirm that the grand canonical determinant is much smaller for
high temperature gauge configurations with complex phases of the Polyakov
loop. Thus the fact
that the Polyakov loop comes out real in dynamical simulations may be
understood as a consequence of center symmetry and its breaking.  

Analyzing the distribution of the canonical determinants as a function of the
quark number, we find for low temperatures a Gaussian-like shape which
widens as the temperature is increased. Using a simple model description we 
relate the width of the distribution to the quark number susceptibility at
$\mu = 0$.
 
An important caveat
must be kept in mind: The numerical results were obtained using ensembles from
pure gauge theory and thus a back-reaction of the fermions or of the
chemical potential is not taken into account in this analysis. In particular the 
distribution of the quark sectors will change. Other aspects, such as the 
center properties which are based on geometry (winding classes of loops) 
may be expected to be universal. In order to understand the back-reaction of the 
fermions a study where we 
explore the properties of canonical determinants in configurations from a 
simulation with dynamical fermions is in preparation.  
 
\vskip8mm
\noindent
{\Large Acknowledgments}
\vskip4mm
\noindent
We thank Andrei Alexandrou, Shailesh Chandrasekharan, Philippe de Forcrand,
Sandor Katz, Tamas Kovacs,  Keh-Fei Liu, Stefan Olejnik,  Bernd Jochen
Schaefer and Kim Splittorf  for discussions. This research was supported in
part by the Slovak Grant Agency for Science, Project VEGA No.\ 2/0070/09, by
ERDF OP R$\&$D, Project CE QUTE ITMS NFP 262401022, by the Center of
Excellence SAS QUTE (L.L.), and EU FP7 project HadronPhysics2. The numerical
computations  were performed on the cluster of the Department of Complex
Physical Systems (Institute of Physics, Bratislava), and at the ZID,
University of Graz.


\begin{thebibliography}{1234567}

\bibitem{canonical1}
  A.~Li, X.~Meng, A.~Alexandru, K.~F.~Liu,
  PoS {\bf LATTICE2008} (2008) 178;
%
  A.~Li, A.~Alexandru, K.~F.~Liu,
  PoS {\bf LAT2007} (2007) 203;
%
  A.~Alexandru, A.~Li, K.~F.~Liu,
  PoS {\bf LAT2007} (2007) 167;
%
  A.~Alexandru, M.~Faber, I.~Horvath, K.~F.~Liu,
  Phys.\ Rev.\  D {\bf 72} (2005) 114513;
%
  A.~Li, A.~Alexandru, K.~F.~Liu, X.~Meng,
  Phys.\ Rev.\  D {\bf 82} (2010) 054502.
  
\bibitem{kentuckymethod}
  X.~Meng, A.~Li, A.~Alexandru, K.~F.~Liu,
  PoS {\bf LATTICE2008} (2008) 032.

\bibitem{canonical2}
  P.~de Forcrand, S.~Kratochvila,
  Nucl.\ Phys.\ Proc.\ Suppl.\  {\bf 153} (2006) 62;
%
  PoS {\bf LAT2005} (2006) 167;
%
  Nucl.\ Phys.\ Proc.\ Suppl.\  {\bf 140} (2005) 514.

\bibitem{Ploopcanonical}
  S.~Kratochvila, P.~de Forcrand,
  Phys.\ Rev.\  D {\bf 73} (2006) 114512.

\bibitem{canonical3}
  S.~Ejiri,
  Phys.\ Rev.\  D {\bf 78} (2008) 074507.

\bibitem{DaGa}
  J.~Danzer, C.~Gattringer,
  Phys.\ Rev.\  D {\bf 78} (2008) 114506.

\bibitem{DaGa2}
  J.~Danzer, C.~Gattringer, L.~Liptak,
  PoS {\bf LAT2009} (2009) 185.
  
\bibitem{altfactwilson}
  A.~Alexandru, U.~Wenger,
  arXiv:1009.2197 [hep-lat];
%
  K.~Nagata, A.~Nakamura,
  arXiv:1009.2149 [hep-lat].
  
\bibitem{altfact}
  P.~Gibbs, Phys.~Lett.~B {\bf 172} (1986) 53;
%
  A.~Hasenfratz, D.~Toussaint, Nucl.~Phys.~B {\bf 371} (1992) 539;
%
  A.~Borici,
  Prog.\ Theor.\ Phys.\ Suppl.\  {\bf 153} (2004) 335.

\bibitem{review}
  P.~de Forcrand,
  PoS {\bf LAT2009} (2009) 010.

\bibitem{luweact}
  M.~L\"uscher, P.~Weisz,
  Commun.\ Math.\ Phys.\  {\bf 97} (1985) 59
  [Erratum-ibid.\  {\bf 98} (1985) 433];
  G.~Curci, P.~Menotti, G.~Paffuti,
  Phys.\ Lett.\  B {\bf 130} (1983) 205
  [Erratum-ibid.\  B {\bf 135} (1984) 516].

\bibitem{scale}
  C.~Gattringer, R.~Hoffmann, S.~Schaefer,
  Phys.\ Rev.\  D {\bf 65} (2002) 094503.

\bibitem{tcrit}
  C.~Gattringer, P.~E.~L.~Rakow, A.~Sch\"afer, W.~S\"oldner,
  Phys.\ Rev.\  D {\bf 66} (2002) 054502.

\bibitem{dualcond}
  E.~Bilgici, F.~Bruckmann, C.~Gattringer, C.~Hagen,
  Phys.\ Rev.\  D {\bf 77} (2008) 094007;
%
  arXiv:0812.2895 [hep-lat].

\bibitem{dualcond2}
  F.~Synatschke, A.~Wipf, C.~Wozar,
  Phys.\ Rev.\  D {\bf 75} (2007) 114003;
%
  F.~Synatschke, A.~Wipf, K.~Langfeld,
  Phys.\ Rev.\  D {\bf 77} (2008) 114018.

\bibitem{takahashi}
  T.~T.~Takahashi,
  arXiv:0807.0864 [hep-lat].
 
\end{thebibliography}
\end{document}